# Adapt/Exchange decisions or generic choices: Does framing influence how people integrate qualitatively different risks?

Romy Müller & Alexander Blunk

Chair of Engineering Psychology and Applied Cognitive Research, Faculty of Psychology, TUD Dresden University of Technology, Helmholtzstr. 10, 01069, Dresden, Germany

**Abstract**

In complex systems, decision makers often have to consider qualitatively different risks when choosing between options. Do their strategies of integrating these risks depend on the framing of problem contents? In the present study, participants were either instructed that they were choosing between two ways of solving a complex problem, or between two generic options. The former was framed as a modular plant scenario that required choices between modifying parameter settings in a current module (Adapt) and replacing the module by another one (Exchange). The risk was higher for Adapt to harm the product and for Exchange to harm the plant. These risks were presented as probabilities, and participants were either told that the consequences of both risks were equally severe (content-same group), or that harming the plant was much worse (content-different group). A third group made decisions based on the same probabilities, but received a generic task framing (no-content group). We expected framing to affect risk integration, leading the content-same group to make different choices than the no-content group. Contrary to this hypothesis, these two groups were strikingly similar in their decision outcomes and strategies, but clearly differed from the content-different group. These findings question whether ecological validity can be enhanced merely by framing a task in terms of real-world problem contents.

## Introduction

Psychological decision making studies usually rely on simple choices in minimalistic scenarios. For instance, in probability discounting, participants choose between two options with different gains that differ in a particular risk. This is in stark contrast with the real world, where decision makers often have to integrate multiple, qualitatively different risks. How they perform this integration may have profound personal, social, or environmental consequences, depending on the specific problem context. For instance, consider an industrial operator trying to solve a problem in the production process of a modular plant. This can either be done by modifying parameters in the current plant setup, (e.g., adapt temperature or pressure), or by reconfiguring the physical plant setup (e.g., exchange the reactor module for a bigger one). One solution may come with a higher risk of compromising product quality, while the other may cause damage to the plant. These risks can also differ in the severity of their consequences. For instance, suboptimal product quality might only reduce profits, while a defective plant might also require effortful and costly repair, or put the safety of humans and the environment at stake. Thus, even if the two risks were numerically identical, one of them might have to be weighted more strongly than the other.

How do people weight and integrate different risks? An important influence could be the specific problem content. Previous studies investigated how the framing of problems influences decision making. Structurally identical decisions were either framed as choices in different domains, or choices between abstract, content-free versus concrete, semantically rich options. However, no previous study has investigated how such framing affects the integration of qualitatively different risks. This issue was addressed in the present study. To derive the specific research questions, we first provide a background on framing in decision making and risky choice.

## Do decisions depend on the framing of contents?

In decision making research, problem content has long been treated as irrelevant, assuming that different formulations of structurally identical problems do not alter decisions (*principle of invariance*, Kahneman & Tversky, 1984). While such reductionism certainly has its benefits in terms of experimental control, it has been questioned what the results can teach us about real-world decision making (Goldstein & Weber, 1995). Accordingly, the role of problem content has received increasing attention over the years. It turns out that both decision outcomes and strategies can depend on whether people decide about abstract options, interpersonal relations, medical procedures, or consumer products (e.g., Goldstein & Weber, 1995; Mandel & Vartanian, 2010; Rettinger & Hastie, 2001; Vartanian et al., 2011). Several, not mutually exclusive suggestions about the functional role of problem or domain contents have been put forward (for an overview see Rettinger & Hastie, 2001). For instance, contents may change peoples' familiarity with the problem, the moral relevance of the decision, the personal importance of options, or the expected duration of outcomes. However, this does not always affect decision making. For instance, in a study on the effects of agricultural policies, participants' decisions did not depend on whether the decision context was framed in a concrete or abstract manner (Ferré et al., 2023). Thus, the findings on the effects of problem content are mixed.

## How do problem contents affect risky choice?

Problem contents also influence how people choose between safer and riskier options. For instance, they may change people's risk preferences (Barseghyan et al., 2011; Einav et al., 2012; Pennings & Smidts, 2000; Wölbert & Riedl, 2013) or their valuation of gains and losses (Heilman & Miclea, 2016). On the other hand, domain-general risk attitudes can be a better predictor of risk taking than domain-specific ones across multiple domains (Reeck et al., 2022). The boundary conditions for content effects remain to be clarified. As the present study focuses on strategies of risk integration, it is important that not only people's eventual choices, but also their underlying decision processes and strategies can depend on content (Goldstein & Weber, 1995; Mandel & Vartanian, 2010; Rettinger & Hastie, 2001; Vartanian et al., 2011).

Evidence that different contents induce different mental processes stems from neuroimaging research (Vartanian et al., 2011). Choices between a certain and an uncertain option were either framed as a matter of saving human lives or as monetary gambles. Decisions about life led to higher risk aversion, conflict, and sensitivity to negative feedback. Moreover, they recruited brain areas associated with context-sensitive reward processing. Conversely, gambling led to stronger activations in regions associated with probability signaling and risk prediction. Thus, in life-related decisions, people primarily focused on moral evaluations and meaning, while in the case of gambles they mainly focused on the computations necessary to maximise outcomes.

A broader and more qualitative analysis of content effects on strategy selection was reported by Rettinger and Hastie (2001). Participants chose between a safe and a risky option, and these decisions were framed as being about legal issues, academic grades, stock investments, or gambles. Based on verbal reports, seven strategies were identified: calculation, avoiding the worst outcome, striving for the best outcome, story, moral, feeling, and regret. While the numerical strategies were widely applied across all problem contents, their combination with other strategies varied. For instance, moral considerations played a large role in the legal domain, but not in decisions about academic grades. These strategy shifts were accompanied by shifts in risk preference. At the same time, content did not change the valuation of outcomes, and thus was not reducible to importance effects. Moreover, although participants relied on calculation even in the content-rich domains, the type of calculations was not assessed. Therefore, the results do not allow us to infer how people's strategies of integrating multiple risks will depend on content. The present study addressed this issue in the context of Adapt/Exchange decisions.

## Present study

We aimed to test the assumption that sufficiently complex problem contents are essential for an ecologically valid assessment of decision strategies. In previous studies of Adapt/Exchange decisions, it was argued that the framing as a complex industrial problem might contribute to explaining why participants used satisficing strategies (Müller, 2024; Müller & Pohl, 2023; Müller & Urbas, 2020). In particular, it was speculated that knowing about this complexity kept participants from solely resting their decisions on the explicit numbers representing the costs of Adapt and Exchange. Instead, they might have considered additional, non-numeric features of these options (e.g., efforts and risks).

In the present study, we tested the role of problem content by either providing or not providing such content. Some participants learned about Adapt/Exchange decisions in modular plants before entering the experiment. The scenario was framed as a choice between an option with a higher risk of harming the product (Adapt) and one with a higher risk of harming the plant (Exchange). In the experiment, participants saw four percentages (each risk for each option, see Figure 1), with their magnitudes varying between trials. To test the effects of content framing, we compared choice frequencies and decision strategies between three groups.

The **content-same group** was told that the consequences of both risks were equally severe. This was justified by the product being expensive, and it was stated that the plant could be repaired. Critically, we assumed that this explanation would not suffice to make participants evaluate the risks as equally severe – we expected them to remain cautious not to harm the plant, due to the framing of the scenario as a safety-critical industrial domain. Thus, participants' integration of the two risks should be biased towards plant safety, and thus against choosing Exchange. This hypothesis was investigated by comparing decisions in the content-same group with two other groups. The **content-different group** served as a baseline for a strong bias towards plant safety. The problem was framed as the same Adapt/Exchange scenario, but participants were told that the consequences of harming the plant were much more severe than those of harming the product. As we assumed this to be most people's default, we expected choice patterns in the content-same group to resemble those in the content-different group, with only an overall difference in the frequency of Exchange choices, but similar dependencies on the risk magnitudes and their ratios, and similar strategies of integrating the risks. The **no-content group** served as a baseline for judging the risks as equally severe. Participants were merely instructed that they were choosing between two abstract options (i.e., option 1 and 2) with generic risks (i.e., risk for factor A and B). Note that in the text, we will use the same labels to refer to the options and risks for all three groups to facilitate comparison, but participants in the no-content group were neither aware of the Adapt/Exchange scenario, nor of the risk identities. Thus, no a priori preference for a particular option should exist, but choices should largely depend on a numerical integration of risk values. These choices should clearly differ from the content-same group, if indeed the framing as a complex industrial scenario affected decisions. That is, the content-same group should choose Exchange less often, there should be differences in the dependence of choice frequencies on the risk magnitudes, and differences in strategies.

In sum, we assumed that the framing as a complex industrial scenario would render the instruction of equal risk severity ineffective. Participants should not treat the choice like participants who are unaware of the contents, but more like participants who are instructed in line with the presumed default assumption that harming the plant is more severe.

## Methods

### Participants

Participants were recruited from the university's participant pool and received partial course credit or 7 € per hour. Of the total 66 participants, 21 took part in the no-content group, 23 in the content-same group, and 22 in the content-different group. Participants' age varied between 18 and 63 years ($M = 25.4$, $SD = 7.9$), 43 were female and 23 were male.

### Apparatus and stimuli

**Lab setup** In a quiet lab room, up to three participants worked in parallel. The instructions and main experiment were presented on one of three laptops (13, 14 and 15.6") and a standard computer mouse was used as an input device.

**Instruction** Three instruction videos were based on a Microsoft PowerPoint presentation with audio overlay. They took eleven minutes (content groups) or three minutes (no-content group). Part 1 of the videos for the content groups introduced the modular plant scenario (i.e., application and purpose, two versions of a reactor module, costs and benefits of Adapt and Exchange, two risks). Using specific examples, it was explained what it meant to harm the plant or product, and how these risks differed between Adapt and Exchange. Only one slide about the severity of the two risks differed between the two content groups (i.e., either explaining why both risks were equally severe, or why harming the plant was worse). In Part 2, both content groups received the same explanation of the experimental task, including the presented stimuli and required actions. The video for the no-content group only presented an abstract version of Part 2. It informed participants that they had to choose between two options (option 1 and 2), each of which was associated with two risks (risk for factor A and B).

**Stimuli** The stimuli had a resolution of 1024 × 768 pixels, and each included the following elements (see Figure 1):

(1) Labels of the two solutions (horizontally) and two risks (vertically) in white font on a black background. For the content groups, the labels were "Adapt", "Exchange", "Risk of plant damage", and "Risk of suboptimal product quality". For the no-content group, the labels were "Option 1", "Option 2", "Risk for Factor A", and "Risk for Factor B". All text was presented in German.

(2) A 2 × 2 matrix of boxes containing the risk percentages for all four combinations of options and risks. Percentages were presented in black font (Calibri, 44 pt) on white background, and the boxes for Adapt and Exchange had a green or purple outline, respectively. The numerical values differed between trials, but were identical across groups.

(3) A green button to choose Adapt or Option 1 and a purple button to choose Exchange or Option 2.

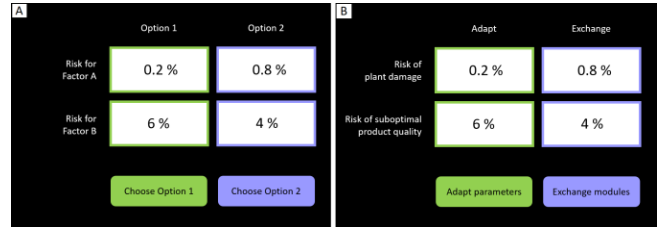

Figure 1: Stimulus example (A) in the no-content group and (B) the two content groups.

The numerical risk percentages varied across trials. Each risk type had four equidistant values, and plant risk (0.2, 0.4, 0.6, or 0.8 %) was ten times smaller than product risk (2, 4, 6, or 8 %). In total, 36 value combinations were presented, with the rule that plant risk always was higher for Exchange than Adapt, while product risk was always higher for Adapt than Exchange. This led to six Adapt/Exchange ratios for plant risk (i.e., 0.2/0.4, 0.2/0.6, 0.2/0.8, 0.4/0.6, 0.4/0.8, and 0.6/0.8) and six for product risk (i.e., 4/2, 6/2, 6/4, 8/2, 8/4, 8/6). Exchange always had a lower total risk than Adapt, as the smallest difference for product risk still exceeded the largest difference for plant risk. Thus, when ignoring risk severity, the numerically optimal solution would be to choose Exchange in each trial.

### Procedure

Before the experiment, participants watched the instruction video and performed a multiple-choice knowledge test about its contents. At least 75 % correct answers were required for inclusion in the data analysis (all participants passed the test). The instruction and test took about 30 minutes.

The experiment used a 3 (*framing*) × 6 (*plant risk A/E ratio*) × 6 (*product risk A/E ratio*) design that varied framing between participants and the two Adapt/Exchange risk ratios within participants. Participants performed one practice block and ten experimental blocks. Each block consisted of 36 trials, corresponding to the 36 combinations of plant and product risk, presented in random order. In each trial, participants saw the four risks (i.e., Adapt-plant, Adapt-product, Exchange-plant, Exchange-product). They had to choose either Adapt or Exchange by clicking the respective button. After this choice, the numbers disappeared, and only the background elements (i.e., text, boxes, buttons) remained on the screen. After 500 ms, the next trial started.

After the experiment, a structured interview was conducted to elicit participants' decision strategies. Participants first answered general questions about their approach of solving the task, and were then shown six example stimuli on paper. For each stimulus, they had to indicate which option they would choose, and then provide a detailed account of how they had proceeded to make this choice. Altogether, the experiment took about one hour.

## Results

### Choice outcomes

The mean percentage of Exchange choices was analysed with a 3 (*framing*) × 6 (*plant risk A/E ratio*) × 6 (*product risk A/E ratio*) mixed-measures ANOVA. Only effects that include framing are reported. All pairwise comparisons were performed with Bonferroni correction.

A main effect of framing, $F(2,63) = 49.149$, $p < .001$, $\eta_p^2 = .609$, indicated that Exchange was chosen less often in the content-different group (31.2 %) than in the content-same and no-content group (75.2 and 75.8 %), both $p$s < .001, while the latter two groups did not differ from each other, $p > .9$.

Framing interacted with plant risk, $F(10,315) = 4.956$, $p < .001$, $\eta_p^2 = .136$ (see Figure 2A). On the one hand, the pattern reported for the main effect was observed for each ratio: less Exchange for the content-different group than the other groups, all $p$s < .001, and no differences between them, all $p$s > .7. On the other hand, while the descriptive pattern of ratio comparisons was the same in all groups, the dependency of choices on plant risk ratio was weakest in the no-content group, where fewer differences between individual ratios were significant (6 out of 15, $p$s < .03) than in the two content groups (10 out of 15 for each, $p$s < .003).

There also was an interaction of framing and product risk, $F(10,340) = 3.234$, $p < .001$, $\eta_p^2 = .093$ (see Figure 2B). Again, the content-different group chose Exchange less often than the two other groups for each ratio, all $p$s < .001, while the content-same and no-content group did not differ for any ratio, all $p$s > .6. Similar to the results for plant risk, choices were least dependent on product risk ratio in the no-content group, where the number of significant differences between ratios was slightly lower (8 out of 15, $p$s < .04) than in the two content groups (11 out of 15 for each, $p$s < .02).

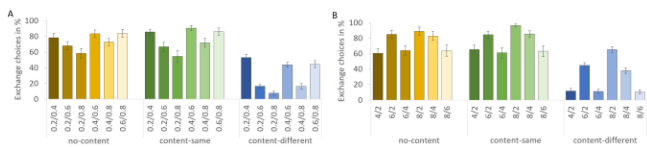

Figure 2: Percentages of Exchange choices in the three groups, depending on the Adapt/Exchange ratio of (A) plant risk and (B) product risk. Error bars represent standard errors of the mean.

Finally, there was a significant three-way interaction of framing, plant risk, and product risk, $F(50,1575) = 9.101$, $p < .001$, $\eta_p^2 = .224$. Put simply, the groups differed as to whether and when plant risk affected decisions (see Figure 3). For the no-content group, plant risk had a noticeable impact when the Adapt/Exchange difference in product risk was low (i.e., 4/2, 6/4, 8/6), whereas plant risk hardly mattered when the Adapt/Exchange difference in product risk was high (i.e., 8/2, 8/4, 6/2). In the latter case, Exchange was highly and consistently preferred (see purple lines in Figure 3A). The content-different group showed the opposite pattern: plant risk had a large impact on Exchange choices when the Adapt/Exchange difference in product risk was high, but not when it was low. In the latter case, Exchange was consistently dispreferred (see green lines in Figure 3C). For the content-same group, the pattern of Exchange choices lay in between the two other groups: plant risk affected choices across the entire range of product risks, except for the highest Adapt/Exchange difference in product risk (i.e., 8/2, see dark purple line in Figure 3B). Thus, framing affected when and how participants integrated the two risks.

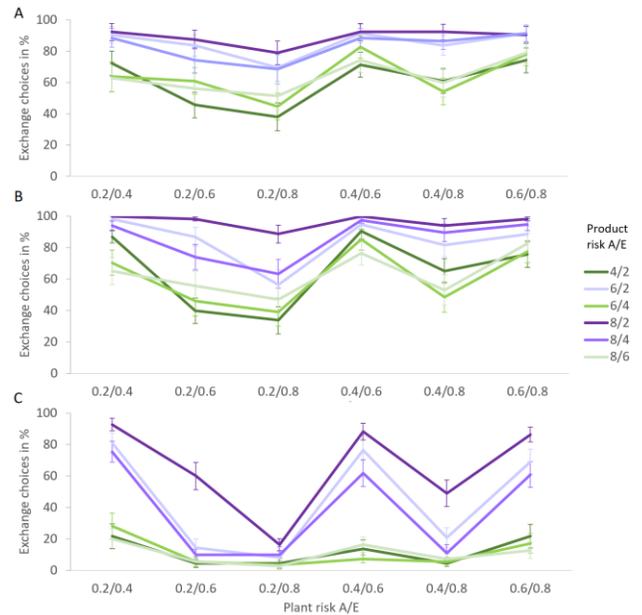

Figure 3: Percentage of Exchange choices depending on the Adapt/Exchange ratios of plant and product risk, for (A) the no-content group, (B) the content-same group, and (C) the content-different group. Error bars represent standard errors of the mean.

### Decision strategies

To investigate participants' strategies of risk integration, we first collected all strategies reported in the post-experimental interview. These reports concerned the relative importance of the two risks, exclusively choosing one option, avoiding particular values, setting decision boundaries, (dis)using the decimal indicator, relying on risk ratios versus differences, and changing strategies during the experiment. We extracted 17 strategies and compared their frequencies between groups. In the following, we will only highlight the most notable differences. All participants in the content-different group considered plant risk to be more important, while only 17.4 % in the content-same group considered both risks to be equally important. Instead, 73.9 % indicated that product risk was more important, seemingly using magnitude as a proxy for importance. In line with this, most participants in the content-different group (77.3 %) reported to have ignored the decimal indicator (e.g., 0.6 % plant risk is equivalent to 6 % product risk). Conversely, this strategy was reported by only 55.2 and 33.3 % in the content-same and no-content groups. While risks of 0.8 or 8 % were avoided by several participants in the content-same and content-different group (34.8 % and 22.7 %), not a single participant mentioned this strategy in the no-content group. Taken together, the reports suggest that some strategies depended on framing. However, they only reflect what participants mentioned spontaneously, while other strategies may have been missed, for instance when they were hard to verbalise or remember.

Thus, to go beyond subjective reports, we checked how consistent participants' actual choices were with a number of strategies. For a given strategy and risk value combination, you can predict how participants should choose, if they were fully adhering to this strategy. For instance, when the strategy is to avoid plant risks of 0.8 %, the percentage of Exchange choices should be 0 % in trials with this value. Any choice frequency above 0 % can be interpreted as a deviation from the strategy. Our aim was to analyse the magnitude of such deviations. To this end, we first defined a set of strategy components based in the subjective reports (see Table 1).

Table 1: Components of strategies with explanations and examples. Combinations of these components were used to describe choice behaviour as reflected in Figure 4.

| Strategy | Explanation and examples |
| --- | --- |
| Abs | Use absolute numbers (i.e., consider decimal indicator), always choose Exchange; Example: Difference between plant risks of 0.2 and 0.8 is smaller than between product risks of 4 and 2 |
| Ratio | Compute option ratio for each risk; choose option that is better on risk with larger ratio; Example: 0.2/0.6 differs more than 8/4 (three- vs. twofold increase), 0.4/0.8 is identical to 4/2 (both times twofold increase) |
| Diff | Compute option difference for each risk; choose option that is better in the risk with the larger difference; Example: 0.2/0.6 is identical to 8/4 (both increase by four units), whereas 0.4/0.8 differs more than 4/2 (increase by four vs. two units) |
| 0.8 | Avoid Exchange when plant risk is 0.8 |
| 8 | Avoid Adapt when product risk is 8 |
| A | If options are identical in all relevant criteria, choose Adapt |
| E | If options are identical in all relevant criteria, choose Exchange |
| R | If options are identical in all relevant criteria, choose randomly |

We combined these strategy components in all coherent ways (i.e., only omitting contradictions), which led to 25 combinations: 1 absolute number strategy, 12 difference strategies, and 12 ratio strategies. For each strategy and trial type, we computed normative choice frequencies. That is, we defined in what percentage of trials Exchange should be chosen when fully adhering to this strategy (i.e., 0, 50 or 100 %). We then computed for each participant by how much their percentage of Exchange choices in the respective trial type deviated from the normative value for each strategy. By averaging these deviations across all participants in a group, we were able to observe how closely the choices of each group matched the 25 strategies (see Figure 4).

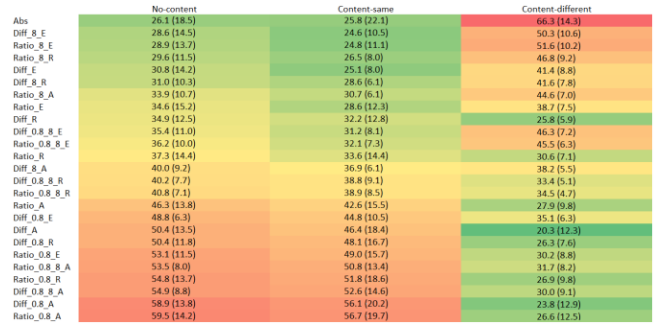

Figure 4: Deviation of choices (*M* and *SD*) from the normative strategy values in all groups, sorted by deviation magnitude in the no-content group. Green colours reflect low and red colours reflect high deviations. Abbreviations correspond to those in Table 1

As can be seen in Figure 4, the patterns of adhering to the strategies were highly similar between the no-content and content-same group. If anything, the content-same group had a slightly higher fit with a wider range of strategies (i.e., more green shading). In contrast, the content-different group clearly differed from both other groups, showing a pattern that was almost the opposite. While the best-fitting strategy in the no-content group (i.e., use absolute numbers, thus always choose Exchange) also had a high fit in the content-same group, it had the lowest fit of all strategies in the content-different group. Here, the best-fitting strategy was to use the differences between options for each risk, and if these differences are equal, choose Adapt. Inconsistencies between the groups did not depend on the use of ratio versus difference strategies. Rather, in the content-same and no-content group, strategies with a good fit often involved avoiding an 8 % product risk and choosing Exchange in case of ambiguity. In the content-different group, strategies with a good fit often involved avoiding a 0.8 % plant risk and choosing Adapt. However, note that all deviations were above 20 %. Exploratory analyses revealed much better fits for individual participants. Apparently, choice behaviour cannot adequately be described as the consistent application of any simple calculation strategy across an entire group.

## Discussion

Does the integration of qualitatively different risks depend on problem content? To answer this question, we framed decisions either as a choice between solutions of a complex industrial problem or as a choice between generic options. Contrary to our hypotheses, when participants with content framing learned that the two risks were equally severe, their decision outcomes and strategies hardly differed from those of participants who worked with generic options. Conversely, their choices and strategies were clearly distinct from those of participants who had learned that one risk was more severe than the other. Thus, domain content per se did not seem to have much impact. These results cast doubts on whether the ecological validity of decision making experiments can be enhanced by using more meaningful content. When the aim is to investigate how people decide about complex issues, it may not be enough to perform simplistic lab experiments and just try to induce complexity via a cover story.

## Why were there no strong effects of framing?

The absence of framing effects in the present study raises the question why previous research did find ample evidence for such effects on decision making in general and on risky choice in particular (Goldstein & Weber, 1995; Mandel & Vartanian, 2010; Rettinger & Hastie, 2001; Vartanian et al., 2011). What was different in our study? First, despite the complex problem framing, our stimuli and trial structure were quite simplistic and repetitive. In total, participants had to make 360 choices, each based on four numbers. This may not leave much room for contemplating the complex industrial domain. Second, the operationalisation of content effects might be problematic. Our hypotheses were built around a central assumption: that participants have content-based preconceptions about risk severity, which cannot simply be erased by instruction. This operationalisation is not only rich in untested assumptions, it also builds on the concept of risk importance. This might be critical, as it has been reported that framing does not affect the subjective importance of outcomes (Rettinger & Hastie, 2001). Thus, our theoretical basis should be carefully reconsidered. Third, our two risks strongly differed in magnitude, making Exchange the normatively better choice throughout the experiment. This might have left little room for framing effects, given that people often have problems when meaningful risks are small (Peters et al., 2009). Thus, several factors may have created unfavourable conditions for framing to be effective.

On the other hand, even in previous research, framing effects have not been found consistently. For instance, it did not matter whether decisions were framed as being about policies in a complex agricultural system, or about generic choices (Ferré et al., 2023). Moreover, risky choice in different domains may be shaped by the same domain-general factors, such as overarching individual risk attitudes, which sometimes are a better predictor than domain-specific factors (Reeck et al., 2022). Thus, a lack of evidence for risk-related decisions to be highly domain-specific might be more common than what it seems like at first glance.

There is one aspect of our results that might cautiously be interpreted as evidence for effects of content framing: the subtle group differences in the specificity of risk integration. First, both for product and plant risk, the pairwise comparisons showed that choices varied more with risk ratios in the two content groups than in the no-content group. Second, a similar conclusion can be drawn from the three-way interaction for trials in which the product risks of Adapt and Exchange diverged more strongly (i.e., increasingly favouring Exchange). In this situation, the content-same group took plant risk into consideration more than the no-content group. Third, the strategy analyses suggest a higher diversification of reported strategies between participants in the content-same group, and a wider range of strategies that match participants' actual choices patterns. Thus, meaningful problem contents might lead people to make more differentiated and contextualised choices, rather than just relying on simple calculations to maximise outcomes (Rettinger & Hastie, 2001; Vartanian et al., 2011). Yet, this cannot actually be interpreted as straightforward, conclusive evidence for an influence of content on choice patterns. In the future, dedicated study designs should test whether content indeed affects the situation-specificity of risk integration.

## Limitations and outlook

Perhaps the main limitation of the present study is that we operationalised framing effects as differences between the content-same and no-content group. In a way, our instruction forced the content-same group to do what they did: ignore their potential preconceptions about risk severity in safety-critical domains, and just treat the two risks as equal. We did this to put the existence of framing effects to a critical test. However, perhaps this did not only make it hard but even impossible for framing effects to emerge. This could be tested by running the experiment with a third content group that receives the modular plant framing, but no information about risk severity. We assume that their decisions would be similar to the content-different group. When running the experiment, we considered this trivial, and given that resource limitations required us to select, we decided not to include this condition. In hindsight, it would have been better to do so.

A limitation of our data analysis is its restriction to mean values, while largely ignoring the interesting insights to be gained from individual differences. Such differences in choice behaviour were immense – and seemingly more so in the content groups, in line with the finding that individual differences are a stronger predictor of decisions in content-rich choices than in generic ones (Ferré et al., 2023). Thus, it would be interesting to infer strategies bottom-up, based on participants' actual choices, instead of only top-down. This would certainly yield more complex strategies than the ones we considered. The relatively weak fits between our strategies and participants' actual choices suggests that this is a promising avenue to be explored further.

Also promising is a more conceptual outlook for future research. Instead of focusing all that much on framing as a surface feature, it is worthwhile to consider the problem structures that decision makers encounter in the real world. A nice package may not suffice to enhance ecological validity: regardless of framing, lottery-like, simplistic experiments correlate poorly with experts' risk-related decision making in complex domains (Rommel et al., 2019). However, there is no straightforward way of determining which aspects of a decision are surface features and which ones are deep, structural features (cf. Mandel & Vartanian, 2010). Thus, both empirical and conceptual work is needed to gain more clarity about this in the context of Adapt/Exchange decisions.

## Acknowledgments

We want to thank Leonie Knöppel, Tim Certa, Kevin Harkin, and Sebastian Schubert for support in data collection. Parts of this work were funded by the German Federal Ministry of Education and Research (BMBF, 02K16C070) and the German Research Foundation (DFG, PA 1232/12-3)